# Cross Fusion and Correlation Beamformer for Row-Column Array Based 3D Ultrasound Imaging

**Qiandong Sun[1,2], Rui He[1,2], Shilin Hou[3], Jiyan Dai[3], and Kailiang Xu[1,2,4]**

[1] College of Biomedical Engineering, Fudan University, Shanghai 200438, China

[2] State Key Laboratory of Integrated Chips and Systems, Fudan University, Shanghai 200438, China

[3] Department of Applied Physics, The Hong Kong Polytechnic University, Hong Kong, China

[4] Poda Medical Technology Co., Ltd., Shanghai 200433, China

*Corresponding author: xukl@fudan.edu.cn

**ABSTRACT**

Row column addressed (RCA) transducers present a promising solution for ultrafast volumetric imaging with a reduced channel count and a large field of view. However, RCA-based 3D imaging is fundamentally limited by severe sidelobe artifacts and a low signal-to-noise ratio (SNR), primarily due to weak transmit focusing inherent in RCA based ultrafast imaging strategies. To overcome these challenges, we propose a cross fusion and correlation (CFAC) method that leverages the incoherence of sidelobe artifacts and noise across datasets acquired using orthogonal apertures and multiple steering angle sets. The performance of the proposed method was validated through simulations, in vitro imaging of a multi-purpose ultrasound phantom, and in vivo experiments, and benchmarked against four established techniques: orthogonal plane wave (OPW) imaging, XDoppler method, row-column-specific frame-multiply-and-sum beamforming (RC-FMAS), and coherent factor (CF) imaging. Simulation results demonstrated that CFAC reduced sidelobe levels by 42.0 dB, 38.9 dB, 28.3 dB, and 25.5 dB compared to OPW, XDoppler, RC-FMAS, and CF, respectively. In phantom experiments, CFAC improved the CNR by up to 17.5 dB. Furthermore, in vivo imaging of a rat kidney showed that CFAC enables visualization of a significantly more detailed microvascular network, achieving a CNR improvement of over 25 dB against all benchmarked methods. In conclusion, the proposed CFAC method effectively suppresses sidelobe artifacts and noise in RCA-based imaging under low-SNR conditions, enabling high-contrast 3D visualization while preserving the high frame rate capabilities of ultrafast ultrasound imaging.

**KEY WORDS**

Three-dimensional (3D) imaging

Ultrafast ultrasound

Row-column addressed (RCA)

Ultrasound beamforming

## I. INTRODUCTION

Three-dimensional (3D) ultrasound imaging provides comprehensive volumetric visualization, overcoming a key limitation of traditional two-dimensional (2D) imaging: susceptibility to out-of-plane motion artifacts. Fully addressed 2D matrix array probes are considered optimal for achieving 3D imaging with isotropic resolution, as they allow dynamic focusing in all three dimensions [1-4]. However, the widespread adoption of fully addressed matrix arrays is hindered by the significant increase in the number of required independent electronic channels and the high cost of probe fabrication.

The row-column addressed (RCA) array offers a cost-effective solution for 3D imaging, enabling a high-volume rate and a large field of view [5-9]. An RCA array can be conceptualized as two orthogonal $N$-element linear transducers, each with a large elevational aperture. This architecture allows for a significant reduction in the required number of electronic channels from $N^2$ to $2*N$, which in turn simplifies both the hardware implementation and the demands on data processing. The basic principle of RCA imaging involves transmitting with one array and receiving with the orthogonal array, following by exchanging their roles to perform another transmission- reception. Depending on the desired trade-off between imaging frame rate and spatial resolution, either plane waves transmission [10] or synthetic aperture techniques [11] can be employed. RCA based system have been successfully implemented in various 3D imaging applications, including functional ultrasound imaging [12], vector flow estimation [13], [14], shear wave elastography [15], super resolution imaging [11, 16-20] and nonlinear sound-sheet ultrasound microscopy [21].

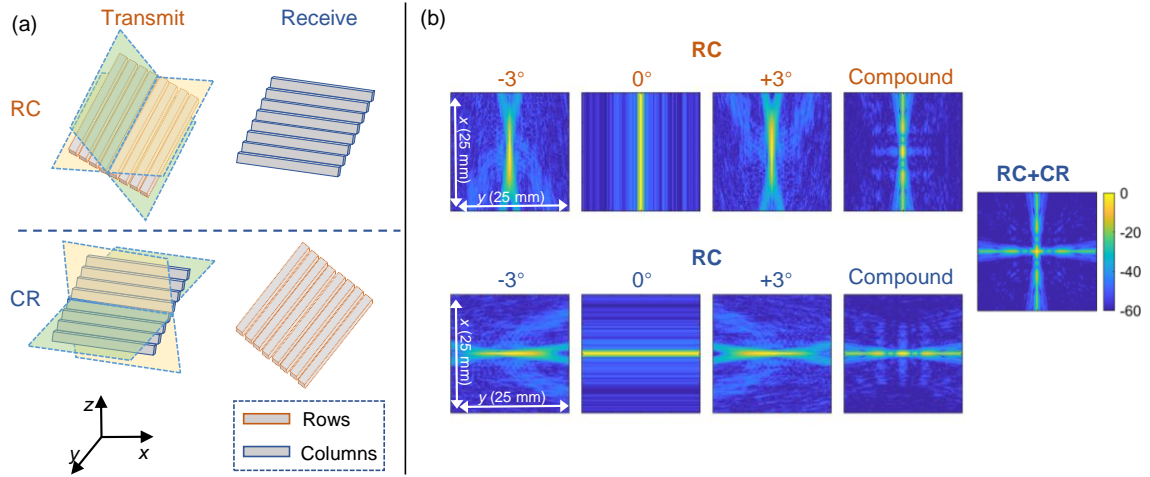

Fig.1. (a) Diagram for the structure of RCA array. (b) PSF images of different individual steering angles and the compounded results. The main lobe is the region where PSFs obtained using different steering angles overlaps.

Although RCA arrays present a promising approach for 3D ultrasound imaging, they are limited by high-level sidelobe artifacts and a low signal-to-noise ratio (SNR), which degrade imaging quality. The inherent weakness in transmit focusing, a consequence of using orthogonal arrays for transmission and reception, leads to significant sidelobes in the elevational direction. Consequently, a single transmit-receive event yields a line-shaped point spread function (PSF), rather than the circular PSF typical of conventional 2D imaging. To recover isotropy, one must swap transmission and reception roles and compound the two orthogonal line-shaped PSFs into a cross-shaped PSF. Multi-angles transmission and then compounding, i.e., orthogonal plane waves (OPW) compounding [10], [22], is developed to reduce the sidelobe level. Nevertheless, a fundamental trade-off exists between volume rate and image quality, as the number of angles that can be used is constrained by the need for ultrafast volumetric acquisition. Moreover, the insufficient transmission focusing also contributes to low contrast. Therefore, developing advanced beamforming strategies or improving transmission focusing that can suppress sidelobes and improve contrast without sacrificing volume rate is essential for advancing RCA-based 3D ultrasound imaging.

Simultaneous azimuth and Fresnel elevation compounding techniques [23], [24], implemented using bias-controlled elements-based RCA arrays, have been developed to improve the transmission focusing. The bias-switchable top-orthogonal-to-bottom electrode (TOBE) [25-27] enables two-way azimuthal focusing, significantly enhancing the imaging quality of RCA. Bertolo *et al.* developed the XDoppler technique [28], which improves RCA-based flow imaging by leveraging spatial decorrelation and the limited spatial overlap of the point spread functions (PSFs) from the two orthogonal apertures. Frame multiply and sum (FMAS) based compounding scheme [29] has also been applied to RCA to suppress the artifacts and increase the contrast. Daya *et al.* [30] introduced an edge-guided, compensated RCA system using multilayered edge-guided stochastically fully connected conditional random fields to improve performance. More recently, Zhang *et al.* [31] developed a spatial-temporal similarity weighting (St-SW) method to further improve the contrast in RCA-based 3D vascular imaging.

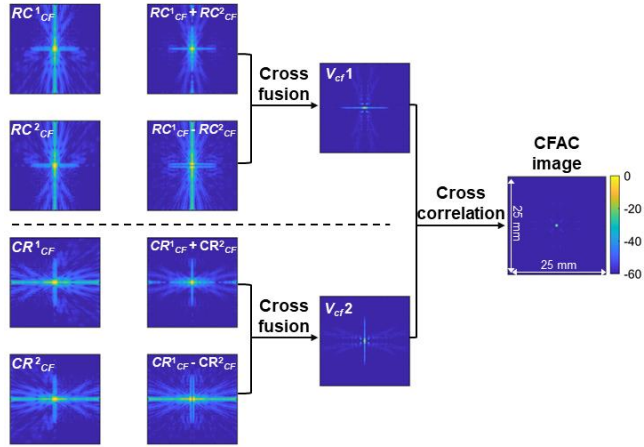

Fig.2. The principle of the cross fusion and correlation method, assuming the RC and CR datasets are split into 2 subgroups.

In this study, we propose a cross fusion and correlation (CFAC) method inspired by Hyper beamforming [32] to enhance imaging contrast in RCA-based 3D ultrasound imaging. The proposed framework performs all operations along the transmit-angle dimension, thereby substantially reducing the computational burden associated with 3D data processing, given that the number of steering angles is considerably smaller than the number of channels. Specifically, a coherent factor (CF) weighting map is first computed for each voxel to suppress sidelobes and noise. This is followed by the application of the CFAC method, which further improves image contrast while maintaining relatively low computational complexity. The effectiveness of the proposed approach is demonstrated through simulations, *in vitro* experiments using a multi-purpose ultrasound phantom, and *in vivo* imaging of the human carotid artery and rat kidney.

## II. Methods

### A. Orthogonal Compounding for RCA

RCA array can be regarded as comprising two orthogonal linear arrays, namely, the row and column arrays [6], as illustrated in Fig. 1(a). Due to the row-column addressing scheme, RCA elements have a large elevation aperture, and focusing can only be performed using a cylindrical law [33]. Consequently, for each plane wave transmission, only receiving focusing is achieved-either by transmitting with the row array and receiving with the column array (RC), or vice versa (CR). There is no transmit focusing in either case.

Although synthetic transmit focusing can be approximated by coherently compounding volumes from multiple steering angles, the need to maintain a high-volume rate restricts the number of transmissions. This limitation results in transmitting focusing that is substantially weaker than the receiving focusing. As a result, the PSF becomes elongated in the transmission direction, as shown in Fig. 1(b). To achieve more isotropic spatial resolution in both the lateral and elevational directions, volumes obtained from both RC and CR acquisitions are coherently compounded—an approach known as orthogonal plane wave (OPW) compounding [10]. Nevertheless, due to the limited number of steering angles, transmitting focusing remains relatively weak, leading to high-level artifacts and a PSF with a pronounced cross-shaped structure.

## B. Coherent Factor (CF)

The coherence factor (CF) is defined as the ratio of coherent energy to incoherent energy, and is expressed as follows:

$$W_{CF} = \frac{\left|\sum_{i=1}^{m} IQ_i(t)\right|^2}{\sum_{i=1}^{m} \left|IQ_i(t)\right|^2},$$

(1)

where $m$ is the number of transmission angles, $IQ_i$ is the beamformed IQ signal from the $i$-th transmission angle. Here, the coherence factor is computed by summing the signals across different steering angles, rather than along the element dimension as in conventional implementations. This modification aims to reduce computational complexity, particularly for 3D image processing, since the number of steering angles is significantly smaller than the number of array elements [34]. However, it inevitably leads to a slight degradation in the effectiveness of the CF. $W_{CF}$ is computed for each imaging pixel and used as an adaptive weight applied to the beamformed signals. The $W_{CF}$ approaches one when steering angle signals are highly phase-coherent, typically corresponding to the main lobe; and approaches zero when the signals are incoherent, as in sidelobe or noise regions. By applying $W_{CF}$ weighting to the beamformer output, the main lobe is selectively enhanced while sidelobe and noise contributions are suppressed, leading to improved image contrast and resolution.

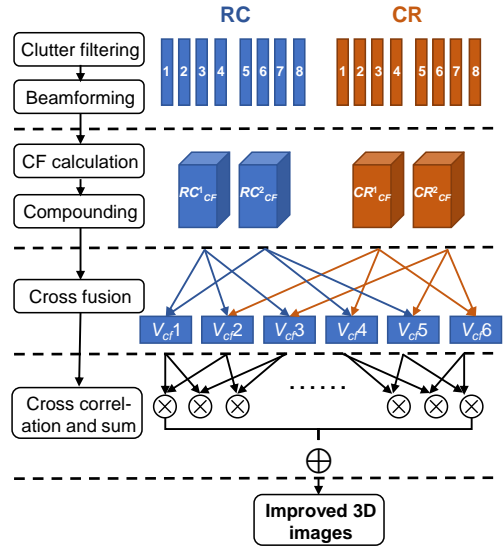

Fig.3. Schematic of the CFAC-2 method. Each frame in this example consists of 8 steered RC transmissions represented by blue blocks and 8 steered CR transmissions represented by brown blocks. The numbers inside the blocks denote the transmission indices. RC and CR datasets are split into 2 subgroups to perform the cross-correlation and sum processing.

## C. Cross Fusion and Correlation

The receiving signals from two orthogonal apertures are first beamformed to obtain the IQ data $RC$ ($x$, $y$, $z$, $m$, $t$) and $CR$ ($x$, $y$, $z$, $m$, $t$), respectively. Here, $x$, $y$, and $z$ represent the lateral, elevational, and axial spatial coordinates of the volumetric data. The index $m \in [1, N_m]$ denotes the steering angle within a single frame comprising $N_m$ steered plane wave transmissions, and $t \in [1, N_t]$ indicates the temporal index across a total of acquired $N_t$ frames.

Next, both the RC and CR datasets are divided into $N$ groups according to the transmit steering angles. As illustrated in Fig. 1, the beamformed signals obtained from different transmit angles exhibit high spatial coherence within the main-lobe region, whereas the coherence is relatively low in the sidelobe and noise regions. Based on this observation, the CF described in the previous section is first computed and applied to each of the $N$ groups for both RC and CR to initially suppress sidelobe artifacts, resulting in a total of $2N$ compounded datasets. Consider the RC case:

$$RC_{CF} = \sum_{i=1}^{m/N} RC(x,y,z,i,t) * \frac{\left| \sum_{i=1}^{m/N} RC(x,y,z,i,t) \right|^2}{\sum_{i=1}^{m/N} \left| RC(x,y,z,i,t) \right|^2}. \tag{2}$$

Assuming $N = 2$, four datasets are obtained: $RC^1_{CF}$, $RC^2_{CF}$, $CR^1_{CF}$, $CR^2_{CF}$. Due to differences in transmission angles and receiving apertures, any two of these datasets exhibit strong overlap in the main lobe region, whereas the spatial coherence in sidelobe and noise regions remains relatively low, as illustrated in Fig. 2. The processing begins with coherent compounding of the RC datasets: P1 = $RC^1_{CF}$ + $RC^2_{CF}$. A noise-kept signal is then derived by computing the difference: N1 = $RC^1_{CF}$ - $RC^2_{CF}$. In this difference-taken volume N1, the sidelobe and noise regions retain high amplitude, while the main lobe amplitude approaches zero. Subsequently, the envelope signal of N1 is subtracted from P1 to cancel out most of the noise and sidelobes, resulting in a noise-canceling volume. The positive portion of abs(P1) - abs(N1) accurately shows the location of the scatterer and the negative portion represents the absence of the scatterer, therefore the negative values are set to zero. The final volume is obtained as follows:

$$V_{cf} = sign(\text{P1+N1}) \cdot \sqrt{(abs(\text{P1}) - abs(\text{N1}))} \\ \cdot \sqrt{((abs(\text{P1}) - abs(\text{N1})) > 0)}. \tag{3}$$

Here, the complex sign is taken to retain the phase information of the signal and the geometric mean is taken to ensure linearity in amplitude. This processing pipeline for producing a single noise-suppressed volume is referred to as "cross fusion". A similar procedure is applied to the $CR^1_{CF}$ and $CR^2_{CF}$ datasets to obtain $V_{cf}2$. Finally, a cross-correlation operation between $V_{cf}1$ and $V_{cf}2$ is performed to generate the final output volume, which achieves uniform resolution in both the lateral and elevational directions.

In fact, the cross-fusion processing can be performed between any two compounded sub-datasets, such as $RC^1_{CF}$ and $CR^1_{CF}$. As shown in Fig. 3, cross-fusion processing is performed pairwise among the four compounded datasets and a total of six noise-canceling volumes are obtained. Then, the cross-correlation operation is carried out and finally combined to obtain the final volume as follows:

$$V_{CFAC} = \sum_{i=1}^{n-1} \sum_{j=i+1}^{n} sign\left(V_{cf}\left(x,y,z,t,i\right) \cdot V_{cf}^{*}\left(x,y,z,t,j\right)\right)$$
$$\cdot \sqrt{abs\left(V_{cf}\left(x,y,z,t,i\right) \cdot V_{cf}^{*}\left(x,y,z,t,j\right)\right)}. \tag{4}$$

The above expression is applicable for generating B-mode images. For Doppler imaging, a similar approach can be applied using the following formulation:

$$\text{Doppler}_{CFAC} = \sum_{i=1}^{n-1} \sum_{j=i+1}^{n} V_{cf}\left(x,y,z,t,i\right) \cdot V_{cf}^{*}\left(x,y,z,t,j\right). \tag{5}$$

In the above introduction, it is assumed that $N = 2$, according to the CFAC-2 method. Following the same principle, RC and CR datasets can be split into a greater number of subgroups to further improve noise suppression performance, such as $N = 4$ for the CFAC-4 method. These two methods are compared with OPW [10], XDoppler [28], RC-FMAS [29] and CF [35] methods in simulation studies, *in vitro* phantom experiments, and *in vivo* evaluations.

TABLE I. SPECIFICATIONS OF THE RCA PROBES

| Specs | Probe 1 | Probe 2 |
|---|---|---|
| Frequency (MHz) | 6.0 | 13.0 |
| Pitch (mm) | 0.2 | 0.15 |
| Kerf (mm) | 0.025 | 0.02 |
| Bandwidth (%) | 80% | 80% |
| Aperture (mm²) | $25.6 \times 25.6$ | $19.2 \times 19.2$ |

## D. Simulations Setup

Simulations were conducted using Field II [36], [37] to evaluate the effectiveness of the proposed method in suppressing sidelobes and noise. We investigated two distinct phantoms: a sparse point-target phantom and a complex letter phantom. The point-target phantom was designed to analyze the PSF and consisted of two groups of scatterers: eight targets with equal amplitudes randomly positioned in the $z = 30$ mm plane, and five targets spaced uniformly along the central axis from 20 mm to 42 mm in depth. The letter phantom, designed to mimic a more complex structure, was composed of scatterers forming the letters "A", "B", and "C". These structures were centered in the $xy$- plane with a 2 mm thickness in $z$, and were populated by scatterers with a uniform spatial density of $10/\lambda^3$ and normally distributed scattering amplitudes. The transmission scheme consisted of a total of 48 plane waves (24 + 24) with an angular pitch of 0.5°. This angular spacing was calculated using the formula $\Delta\alpha = \lambda / L_{x/y}$ [12] to minimize grating lobes, where $\lambda$ is the wavelength and $L_{x/y}$ is the aperture size of the probe in the lateral or elevational direction. Gaussian noise was added to the channel RF data at an SNR of 10 dB using the *awgn* function in MATLAB. The simulated RCA array used the same specifications as the commercial RCA probe (RC6gV, Vermon, 6.0 MHz).

## E. Experimental Setup

### 1) Multi-purpose ultrasound phantom:

A commercial multi-purpose ultrasound phantom (CIRS, Model 040GSE) was used to evaluated the performance of the presented method in B-mode imaging. The RCA probe (RC6gV, Vermon, 6.0 MHz) was connected to a commercial 256-channel programmable platform (Vantage-256, Verasonics, WA,

USA). The relevant probe parameters are listed as 'Probe 1' in Table I. A 2-cycle sine pulse at 6 MHz was transmitted with a peak voltage of 20 V. The transmission scheme included a total of 48 plane waves (24 + 24) spanning a total angular range of 10°, resulting in a volume acquisition rate of 300 Hz. A total of 10 volumes was collected to obtain the final image.

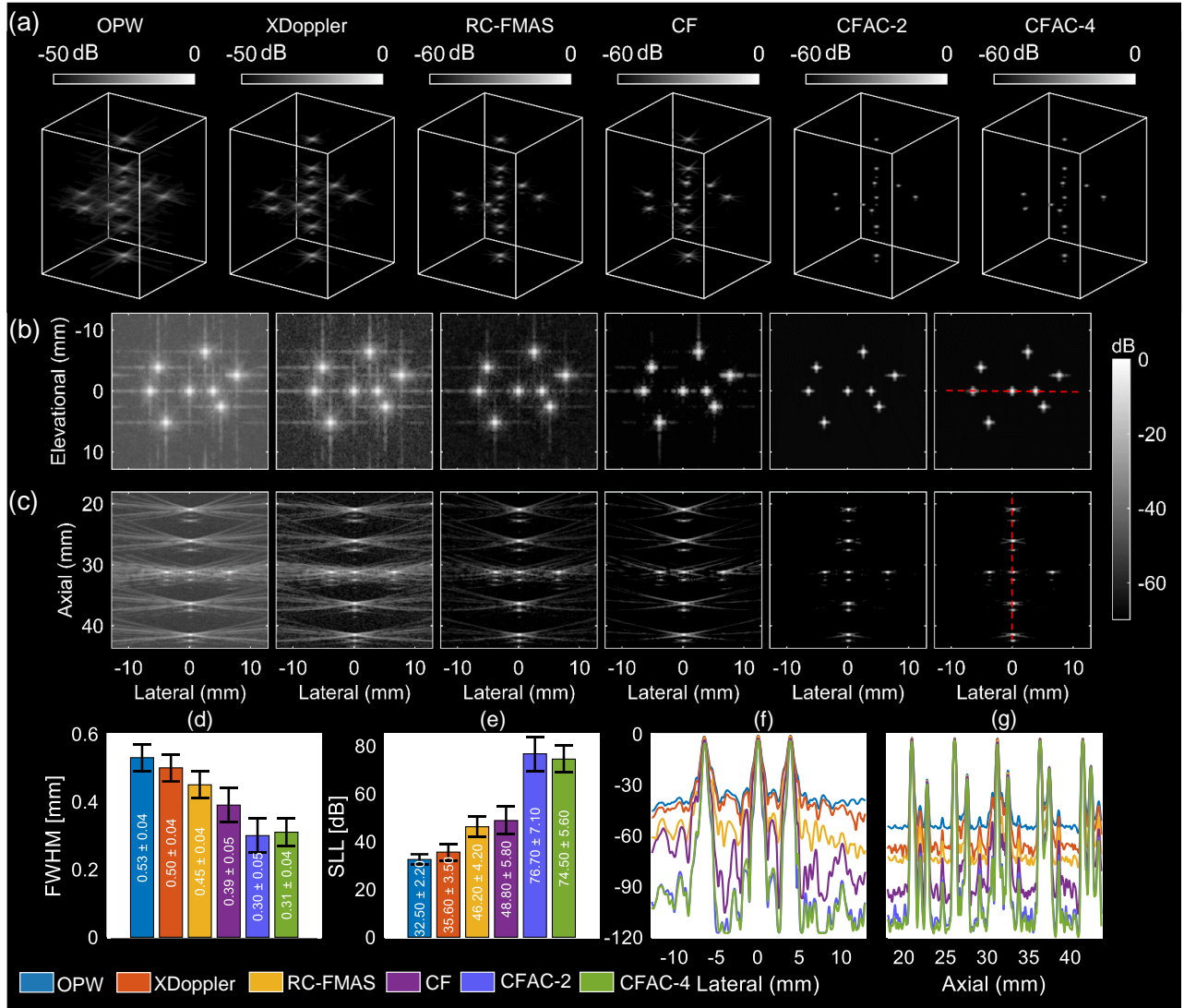

Fig.4. (a) Simulated 3D volumes reconstructed using OPW, XDoppler, RC-FMAS, CF, CFAC-2, and CFAC-4 methods. (b) 2D horizontal slices at a depth of 30 mm. (c) 2D slices at y = 0 mm. Quantitative results of means and standard deviations of FWHM in (d), lateral and elevational SLL in (e) for all the point targets reconstructed using different methods. (f) and (g) correspond to the profiles reconstructed using different methods along the red dashed line in (b) and (c).

## 2）In vivo human carotid artery:

This study was approved by the Human Research Ethics Committee, Huashan Hospital affiliated at Fudan University (2021-065). The parameters of RCA probe and imaging pipeline were consistent with those in the phantom experiment. The ultrasonic sequences complied with the Food and Drug Administration (FDA) recommendations (Track 3) in terms of spatial-peak temporal-average ($I_{SPTA.3}$), and mechanical index (MI). At a transmit voltage of 20 V, the MI and $I_{SPTA.3}$ were measured to be 0.38 and 115.6 mW/cm², respectively. The temperature increase at the surface of the probe was inferior to temperature uncertainty of the measurement system (±0.1°C) during the test sequences. *In vivo* acquisitions were performed on the carotid artery of a healthy volunteer. Prior to the 4D ultrafast acquisitions, a real time B-mode imaging was used to guide probe positioning and confirm visualization of the carotid artery.

### 3）In vivo rat kidney:

All animal experiments were conducted in accordance with the guidelines approved by the Animal Research Ethics Committee of Fudan University (Approval No. 202310031S). The study was performed on two male Sprague-Dawley rats, each aged eight weeks and weighing approximately 300 g. Anesthesia was induced with 5.0% isoflurane in a transparent induction chamber for 5 minutes, after which the animals were placed on a temperature-controlled heating plate to maintain a core body temperature of 37 °C. The isoflurane concentration was then reduced to 1.0% to maintain a stable anesthetic depth while minimizing the risk of adverse side effects. For ultrasound acquisition, the left kidney was externalized through a midline abdominal incision. The exposed organ was positioned on an acoustic absorber and stabilized using two needles inserted into the skin on either side of the kidney, which were then pinned to the absorber. A coupling agent was applied around the periphery of the organ, and a ~5 mm thick acoustic coupling pad was placed between the probe and the kidney to mitigate near-field effects. Key physiological parameters, including body temperature and respiratory rate, were monitored regularly during the acquisition. Microbubble contrast agents were prepared by reconstituting lyophilized SonoVue powder (Bracco, Milan, Italy) with 5 mL of saline, following the manufacturer's instructions. The solution was administered via the jugular vein using a micro-hydraulic infusion pump (LSP01-2B, LONGER) at a constant flow rate of 1.5 μL/s.

Ultrasound acquisitions were performed with a custom-made RCA array probe (RCA13, Poda, Shanghai, China) connected to a commercial 256-channel programmable platform (Vantage-256, Verasonics, WA, USA). The specific probe parameters are summarized under 'Probe 2' in Table I. The imaging sequence employed a total of 32 tilted plane waves (16 + 16) with an angular pitch of 0.5°, transmitted at a center frequency of 13 MHz. This configuration enabled a volumetric frame rate of 500 Hz. A total of 300 volumetric frames were acquired for Doppler imaging.

### F. Data Processing Pipeline

The received RF data were first subjected to IQ demodulation, followed by baseband delay-and-sum (DAS) beamforming implemented in MATLAB R2022b (The MathWorks, Inc., Natick, MA, USA) using GPU acceleration (NVIDIA GeForce RTX 3060 Ti). The voxel size of the reconstructed 3D volume was set to $\lambda/2 \times \lambda/2 \times \lambda/2$ along the $x$, $y$, and $z$ dimensions, respectively. For *in vivo* rat kidney imaging, a singular value decomposition (SVD)-based spatiotemporal filter [38] was applied to the IQ data prior to beamforming to suppress stationary tissue signals and noise. Specifically, for each transmitting plane wave, the after-SVD IQ data were integrated into a 3D matrix of dimension ($n_l$, $n_c$, $m$), where $n_l$ is the RF signal length, $n_c$ is the number of receiving channels, and $m$ is the Doppler ensemble length. This complex-valued matrix was then reshaped into a 2D Casorati matrix of size ($n_l \times n_c$, $m$), and the largest 40 of the 300 singular vectors were discarded to isolate dynamic blood flow signals for subsequent flow visualization.

### G. Performance Metrics

Full width at half maximum (FWHM) was used to quantify spatial resolution. Axial, lateral, and elevational FWHMs were measured from the respective one-dimensional profiles intersecting the PSF at its maximum. The axial, lateral, and elevational FWHMs were measured on the respective profiles intersecting the PSF maximum. The sidelobe level (SLL) was defined as the amplitude of the first sidelobe relative to the main lobe peak, expressed in decibels (dB). To evaluate imaging contrast, the contrast-to-noise ratio (CNR) was computed from the reconstructed images using the following formula:

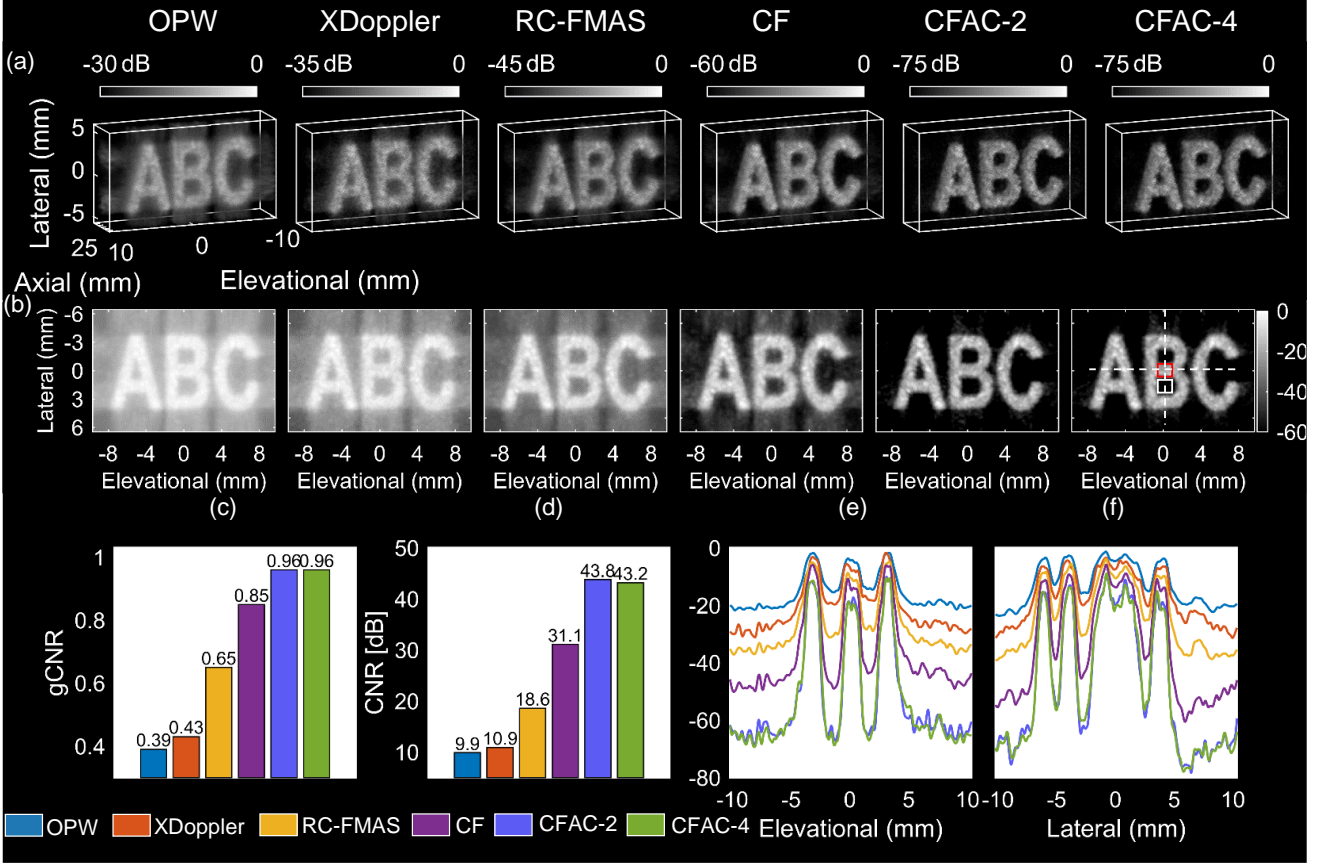

Fig.5. (a) 3D renderings using different methods. (b) 2D horizontal slices at a depth of 25 mm. (c) and (d) display the computed gCNRs and CNRs of the two selected regions, respectively. (e) and (f) show the profiles reconstructed using different methods along the white dashed line in (b).

$$\text{CNR} = 20 \log \left( \frac{\mu_{ROI}}{\mu_{bkgd}} \right), \tag{6}$$

where $\mu_{ROI}$ and $\mu_{bkgd}$ denote the mean signal intensities within the target and background regions, respectively. The generalized contrast-to-noise ratio (gCNR) [39], which is robust to changes in dynamic range and less sensitive to outliers, was used to quantitatively assess contrast. It is defined as:

$$gCNR = 1 - OVL, \tag{7}$$

where $OVL$ denotes the overlap area between the probability density functions (PDFs) of the target and background regions. A lower $OVL$ (and thus a higher gCNR) indicates greater separability between the two distributions, reflecting improved image contrast.

## III. RESULTS

### A. Simulations

Fig. 4(a) presents the simulated 3D rendering volumes reconstructed using various compounding methods,

corresponding to OPW, XDoppler, RC-FMAS, CF, CFAC-2 and CFAC-4. To enable a fair visual comparison, the dynamic range for each method was manually optimized. Fig. 4(b) and Fig. 4(c) depict horizontal slices at a depth of 30 mm and vertical slices at $y = 0$ mm, respectively. These images collectively demonstrate that the proposed CFAC-2 and CFAC-4 methods outperform OPW, XDoppler, RC-FMAS and CF in suppressing sidelobe artifacts and noise.

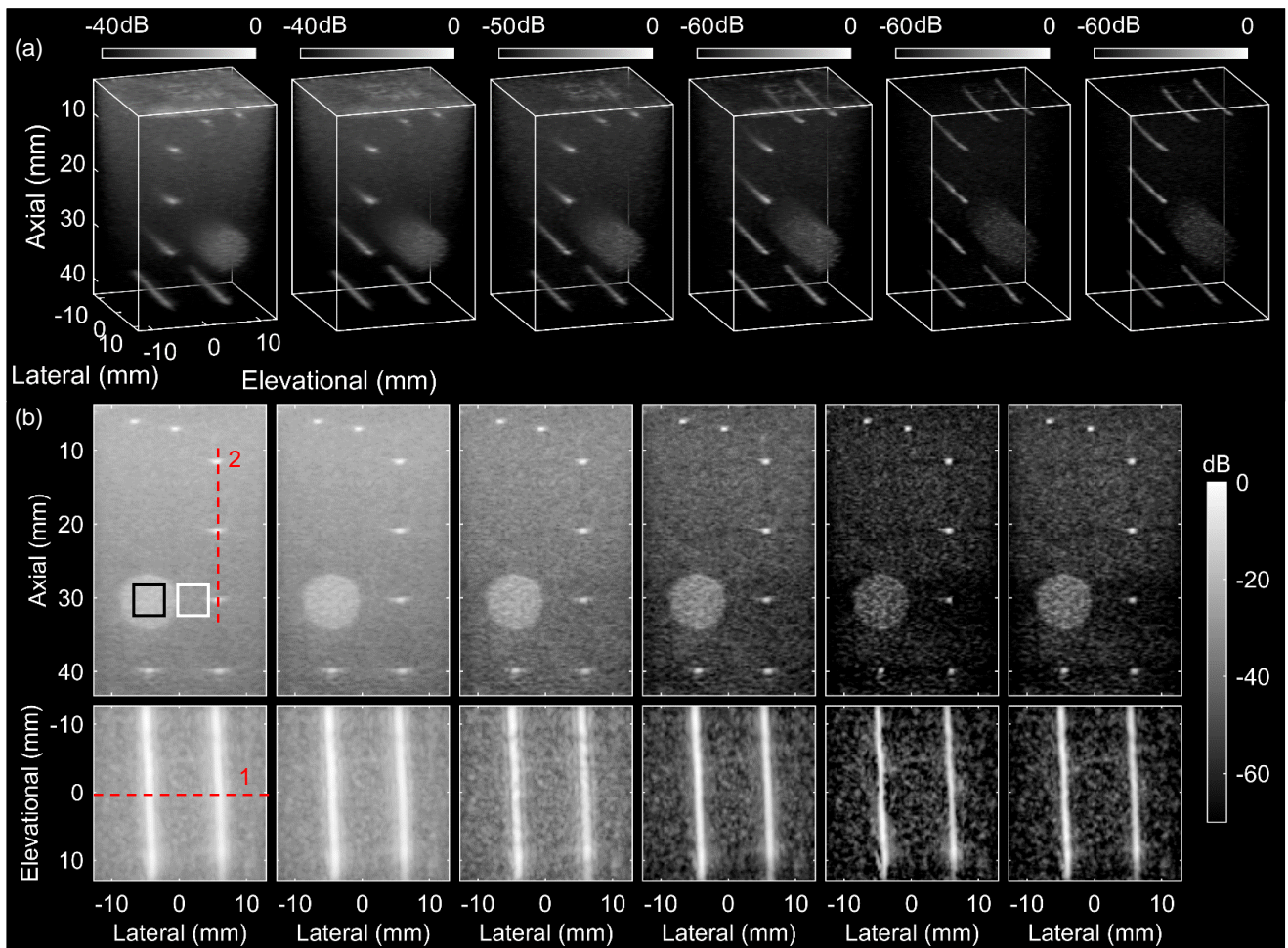

Fig.6. (a) 3D imaging results of the ultrasound phantom. (b) 2D slices (top: at $x = 0$ mm; bottom: horizontal slices of the two hyperechoic lines.

Fig. 4(d) quantifies the lateral FWHM, while Fig. 4(e) presents the measured sidelobe levels in both the lateral and elevational directions. The proposed CFAC method achieves a significant improvement in spatial resolution, with lateral FWHMs of 0.30 mm and 0.31 mm for CFAC-2 and CFAC-4, respectively. In contrast, the FWHMs obtained by OPW, XDoppler, RC-FMAS and CF are 0.53 mm, 0.50 mm, 0.45 mm and 0.39 mm, respectively. In terms of sidelobe suppression, OPW performs the worst, exhibiting a sidelobe level of -32.5 dB, whereas CFAC-2 and CFAC-4 achieve the best performance, reaching -76.7 dB and -75.4 dB, respectively. Fig. 4(f) displays the lateral profiles extracted along the red dashed line in Fig. 4(b), respectively, visually demonstrating the effectiveness of the proposed method in suppressing sidelobe artifacts. Fig. 4(g) shows the axial profile along the dash line in Fig. 4(c), further illustrating the improvement in imaging contrast achieved by the proposed approach.

Fig. 5 presents the simulation results of the letter-phantom experiment. Fig. 5(a) shows the 3D renderings obtained using different methods, each displayed with manually adjusted dynamic ranges for visual clarity. Fig. 5(b) illustrates the corresponding 2D axial slices, where the proposed method demonstrates a marked improvement in imaging contrast. The reconstructed images using the proposed method presents a notable reduction in sidelobe artifacts, particularly around high-scattering regions. Fig. 5(c) and Fig. 5(d) show the computed gCNR and CNR, respectively, evaluated within two regions of interest: a high-scattering region (red box) and a background region (white box) as indicated in Fig. 5(b). The proposed CFAC-2 and CFAC-4 methods achieved superior gCNR values of 0.96, outperforming OPW (0.39), XDoppler (0.43), RC-FMAS (0.65), and CF (0.85). Similarly, in terms of CNR, the CFAC method yielded improvements of 33.9dB, 32.9dB, 25.2dB, and 12.6dB over OPW, XDoppler, RC-FMAS, and CF, respectively. Fig. 5(e) and Fig. 5(f) display the intensity profile along the white dashed line in Fig. 5(b). While the proposed method introduces a slight reduction in the amplitude of high-scattering regions, the intensity ratio between the high-scattering target and background remains superior compared to other methods. Specifically, the relative intensity differences are -18.3 dB, -23.1 dB, -28.8 dB, -39.3 dB, -52.6 dB, and -52.0 dB for OPW, XDoppler, RC-FMAS, CF, CFAC-2, and CFAC-4, respectively.

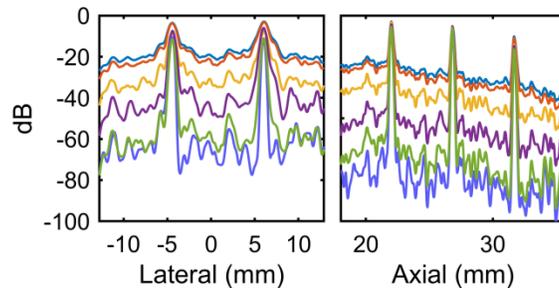

Fig.7 shows the profiles reconstructed using different methods along the red dashed line in Fig. 6(b).

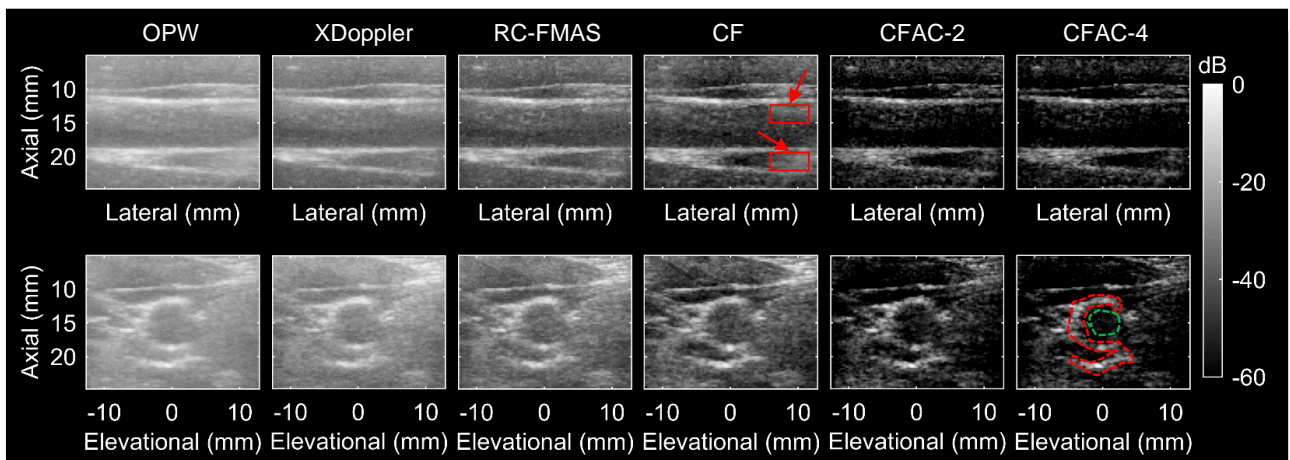

Fig.8. The slices of human carotid artery reconstructed using different methods. The red arrows highlight typical sidelobe artifacts.

## B. Multi-purpose Ultrasound Phantom

Fig. 6(a) illustrates the 3D imaging results of the multi-purpose ultrasound phantom. The proposed method exhibits a significant reduction in background noise and an enhanced contrast of hyperechoic targets. In particular, it effectively suppresses strong near-field noise, enabling clearer reconstruction of hyperechoic lines. Notably, the hyperechoic lines reconstructed using CFAC-4 appear smoother and more continuous compared to those obtained with CFAC-2. Fig. 6(b) presents axial slices extracted from two representative regions: the upper row shows slices along the lateral center of the phantom, while the lower row corresponds to axial slices intersecting the two hyperechoic lines at the bottom. These images demonstrate that the proposed method, especially CFAC-4, successfully suppresses background noise while preserving speckle patterns within the hyperechoic regions. Additionally, for the two hyperechoic lines, the method significantly reduces sidelobe artifacts and enhances edge clarity.

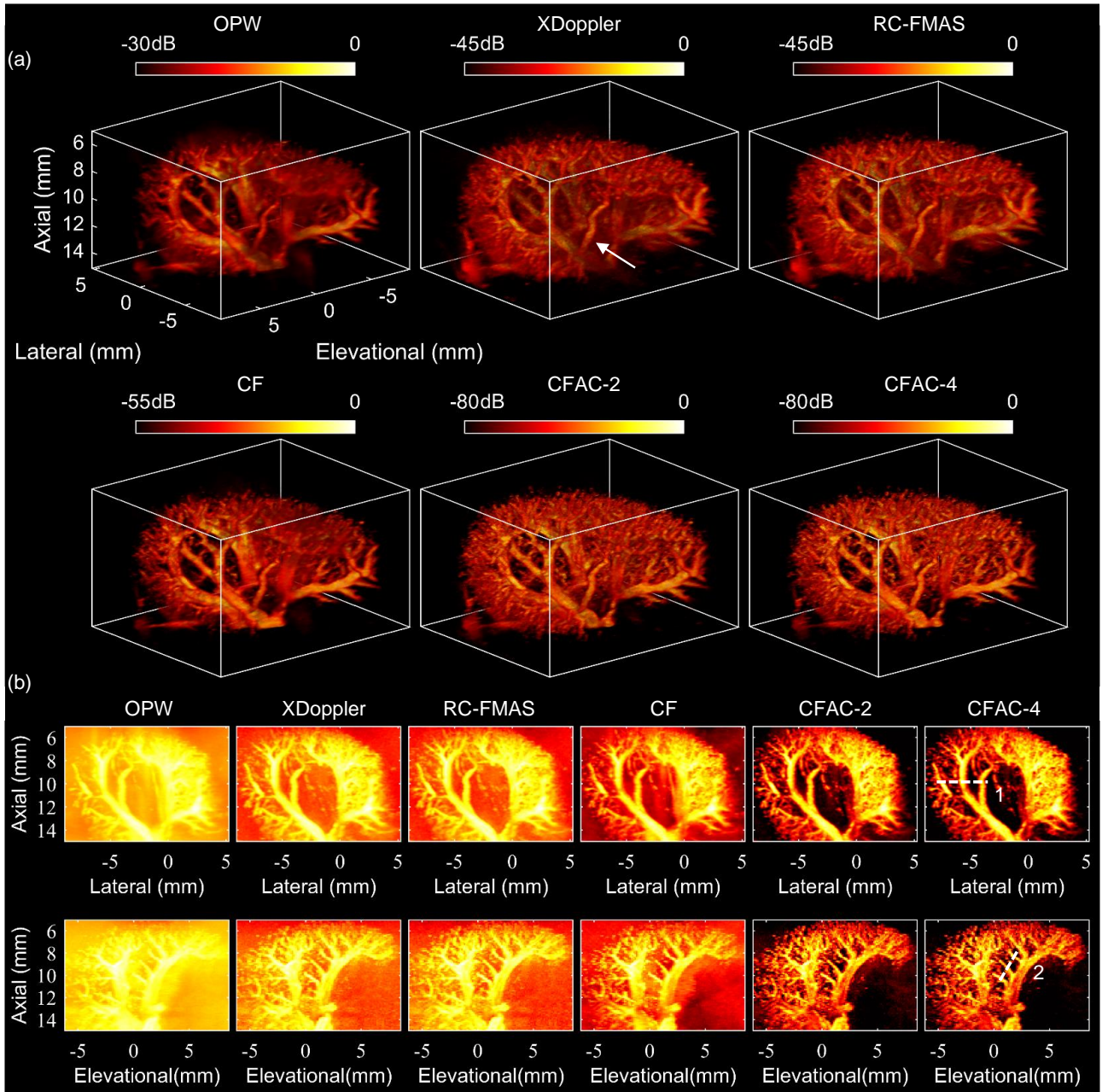

Fig.9. (a) 3D *in vivo* rat kidney volumes reconstructed using OPW, XDoppler, RC-FMAS, Cf, CFAC-2, and CFAC-4 methods. An integer dynamic range is manually selected for each image to ensure clear visualization of the microvasculature without being obscured by background noise. (b) and (c) are the maximum intensity projection slices along the elevational and lateral directions, respectively, each encompassing a projection thickness of 5 mm.

Fig. 7 shows the intensity profiles along the red dashed lines in Fig. 6(b), with the left and right panels corresponding to line-1 and line-2, respectively. The calculated CNR and gCNR for the two selected regions, where the black box denotes the hyperechoic target and the white box denotes the background, are detailed in

Table II. The proposed method yielded the highest average value of 28.7 dB, compared to 11.4 dB, 12.7 dB, 17.3 dB, and 21.2 dB for OPW, XDoppler, RC-FMAS, and CF, respectively. For the gCNR, the proposed CFAC-2 and CFAC-4 methods achieved superior gCNR values of 0.84 and 0.85, respectively, outperforming OPW (0.39), XDoppler (0.43), RC-FMAS (0.65), and CF (0.85).

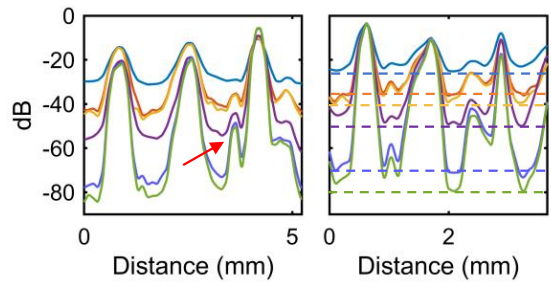

Fig.10. Intensity profiles corresponding to the white dotted lines in Fig. 9(b). Red arrow indicates small vessels that are detectable in XDoppler, RC-FMAS, CF, CFAC-2, and CFAC-4 images, but not in OPW images.

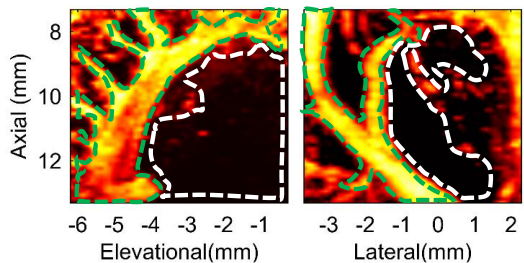

Fig.11. Zoomed-in view of the 2D slice reconstructed using the CFAC-4 method. The green and white dashed lines indicate the manually selected vessel regions and background regions, respectively.

TABLE II. THE COMPUTED CNR AND gCNR OF VALIDATION EXPERIMENTS

| | | CNR (dB) & gCNR | | |
|---|---|---|---|---|
| | Phantom | Human carotid | Rat kidney-1 | Rat kidney-2 |
| OPW | 11.4 & 0.45 | 19.4 & 0.70 | 11.4 & 0.45 | 19.4 & 0.66 |
| XDoppler | 12.7 & 0.49 | 20.3 & 0.72 | 18.3 & 0.65 | 26.9 & 0.82 |
| RC-FMAS | 17.3 & 0.62 | 24.2 & 0.77 | 19.0 & 0.67 | 27.5 & 0.83 |
| CF | 21.2 & 0.72 | 28.7 & 0.85 | 24.8 & 0.79 | 38.6 & 0.93 |
| CFAC-2 | 28.6 & 0.84 | 34.4 & 0.91 | 45.3 & 0.96 | 53.1 & 0.98 |
| CFAC-4 | 28.9 & 0.85 | 35.9 & 0.92 | 49.0 & 0.97 | 56.6 & 0.98 |

## C. In Vivo Human Carotid Artery

Fig. 8 shows the elevational and axial slices of in vivo human carotid artery images reconstructed using various methods. The proposed method demonstrates superior performance in contrast enhancement. In particular, images reconstructed with OPW and CF exhibit noticeable artifacts within the two red-marked regions indicated by arrows, whereas the CFAC method effectively suppresses these artifacts. While XDoppler and RC-FMAS partially mitigate grating lobe artifacts, they still produce images with relatively low contrast. Table II presents the computed gCNR and CNR for two selected regions: the hyperechoic boundary of the carotid artery (red dashed line) and the background (green dashed line), as indicated in Fig. 8. The proposed CFAC-2 and CFAC-4 methods achieved superior gCNR values of 0.91 and 0.92, respectively, outperforming OPW (0.70), XDoppler (0.72), RC-FMAS (0.77), and CF (0.85). Similarly, in terms of CNR, the CFAC methods exhibited substantial improvements, with average gains of 15.9 dB, 14.8 dB, 10.7 dB, and 6.9 dB over OPW, XDoppler, RC-FMAS, and CF, respectively.

**D. In Vivo Rat Kidney**

Fig. 9(a) presents 3D volumes reconstructed using OPW, XDoppler, RC-FMAS, CF, CFAC-2, and CFAC-4 methods, each rendered with individually optimized dynamic ranges to facilitate qualitative comparison. The proposed method demonstrates superior noise suppression, enabling clearer visualization of vascular structures without obscuration and with improved dynamic range. Figs. 9(b) displays 2D maximum intensity projection slices along the elevational and lateral directions, each encompassing a projection thickness of 5 mm. All slices are visualized using the same dynamic range to emphasize the effectiveness of the proposed methods in suppressing background noise.

Fig. 10 presents two groups of intensity profiles extracted along the white dashed lines shown in Figs. 9(b). Red arrows indicate small vessels that are visible in the XDoppler, RC-FMAS, CF, and CFAC reconstructions but are obscured by noise in the OPW results. Importantly, the contrast of these vessels is notably higher with the proposed CFAC methods. For example, in Profile 2, the background levels, defined as the difference between the peak intensity and background intensity, are 57 dB and 66 dB for CFAC-2 and CFAC-4, respectively, compared to 13 dB, 28 dB, 31 dB, and 36 dB achieved by OPW, XDoppler, RC-FMAS, and CF.

Fig. 11 presents the zoomed-in views of a 2D slice reconstructed using the proposed method. Regions of interest (ROIs) corresponding to vessels and background are manually delineated by green and white dotted lines, respectively. These ROIs are used for the quantitative evaluation of CNR and gCNR, as illustrated in Table II. The results demonstrate that the proposed method yields a substantial improvement in CNR and a notable enhancement in gCNR, thereby validating its effectiveness in improving image contrast.

## IV. DISCUSSION

In this paper, we proposed a novel Cross Fusion and Correlation (CFAC) method for RCA based 3D ultrasound imaging. This approach was designed to mitigate severe sidelobe artifacts and enhance image contrast, which are critical issues stemming from weak transmit focusing and insufficient acoustic energy density. The efficacy of the proposed method was systematically evaluated through numerical simulations, *in vitro* phantom experiments, and *in vivo* imaging. Our results consistently demonstrate that CFAC outperforms existing techniques, including OPW, XDoppler, RC-FMAS, and CF, in terms of both sidelobe suppression and contrast enhancement.

The proposed framework, along with the existing techniques used for comparison, performs all operations along the transmit-angle dimension. This approach significantly reduces the computational burden associated with 3D data processing, as the number of steering angles is much smaller than the number of channels. Consequently, in this work, we exclude methods that operate along the channel dimension, such as the CF-FMAS combination [40], from our comparison, as this would require one of the methods to be executed along the channel dimension.

In simulations, the proposed CFAC-4 method demonstrated a significant reduction in sidelobe levels by 42.0 dB, 38.9 dB, 28.3 dB, and 25.5 dB compared to OPW, XDoppler, RC-FMAS, and CF, respectively. Furthermore, improved lateral FWHM resolutions of 0.30 mm and 0.31 mm were achieved with CFAC-2 and CFAC-4, respectively, outperforming the corresponding values obtained with OPW (0.53 mm), XDoppler (0.50 mm), RC-FMAS (0.45 mm), and CF (0.39 mm). In the multi-purpose ultrasound phantom experiment, the CFAC-4 method enhanced the CNR of the hyperechoic region by 17.5 dB, 16.2 dB, 11.6 dB, and 7.7 dB relative to OPW, XDoppler, RC-FMAS, and CF, respectively. It can be observed that, compared to CFAC-2, CFAC-4 better preserves speckle characteristics in high-echo regions, as shown in Fig. 6. This improvement is likely due to the involvement of more sub-datasets in the cross-correlation and summing process, which helps compensate for some of the amplitude loss. Additionally, this issue can be mitigated by coherently compounding multi-frame. The 3D renderings further confirmed that CFAC effectively suppressed near-field noise artifacts. It is worth noting that CF-based beamforming may cause speckle pattern distortion, particularly near cyst boundaries. This occurs because the adaptive weighting suppresses incoherent echoes, which can inadvertently reduce speckle randomness and local texture fidelity. As the adaptivity increases (e.g., in CFAC-2 and CFAC-4), the effect becomes more pronounced. Nevertheless, this trade-off is accompanied by enhanced cyst contrast and clutter suppression.

In the human carotid artery evaluation, CFAC effectively eliminated sidelobe artifacts near the vessel walls, achieving a high CNR of 35.9 dB, compared to 19.4 dB, 20.3 dB, 24.2 dB, and 28.7 dB for OPW, XDoppler, RC-FMAS, and CF, respectively. In the *in vivo* rat kidney imaging, the proposed method not only reconstructed a more complete microvascular structure but also improved the CNR by over 25 dB compared to the other techniques. The method demonstrated enhanced performance in Doppler imaging, likely due to the point scattering properties of microbubbles, though further investigation is needed to fully understand this effect.

Compared to the OPW, XDoppler, and CF methods, the proposed CFAC approach provides a substantial improvement in image contrast for a modest increase in computational cost. Notably, CFAC is more computationally efficient than RC-FMAS. This advantage stems from our approach of first compounding all steering angles before applying the cross-correlation, whereas RC-FMAS performs this computationally intensive step for each individual angle. The performance of CFAC improves as the number of transmission angles increases, though this enhancement comes at the expense of the volume rate. Conversely, using too few transmission angles degrades performance, as it reduces the spatial coherence between the angle-compounded sub-datasets. This establishes a clear trade-off between image quality and frame rate. A similar trade-off exists internally between performance and complexity: CFAC-4 yields superior results to CFAC-2 but at a higher computational cost. Therefore, the number of sub-dataset groups ($N$) must be carefully selected to balance image quality enhancements with the overall computational load.

The primary limitation of the CFAC framework is its sensitivity to temporal decorrelation, particularly in the presence of rapid tissue motion or high-velocity blood flow. Such motion compromises the spatial coherence between sub-datasets, which can lead to misalignment of the main lobes during cross-correlation and, consequently, to signal suppression. Addressing this limitation constitutes a key direction for our future research. This work will proceed along two main fronts: first, a comprehensive characterization of the algorithm's performance under various motion conditions, and second, the development and integration of a motion compensation strategy, which has proven effective in 2D imaging before compounding [41].

This study evaluates the performance of the proposed method under the conventional plane wave transmission scheme, which is limited to a rectilinear field of view in front of the RCA transducer. Several studies have investigated the potential for expanding the acoustic energy distribution by employing a double-curved diverging acoustic lens placed on top of the RCA, or using a curved toroidal RCA [42], in order to achieve a larger field of view [18]. The proposed method, in principle, could be effectively applied to these techniques, potentially enhancing their applicability for RCA-based cardiac imaging.

The proposed CFAC method also shows significant potential for application in RCA-based ultrasound localization microscopy (ULM). By effectively suppressing sidelobe artifacts, CFAC can substantially improve the isolation and localization of individual microbubbles, a critical challenge in scenarios with high microbubble concentrations. This enhanced localization accuracy, in turn, can reduce the total acquisition time required to reconstruct a volumetric super-resolution image. Such an improvement would be highly beneficial for accelerating fast or high-throughput ULM applications.

The performance of the proposed method was systematically validated through a comprehensive suite of simulations, *in vitro* experiments, and *in vivo* imaging, which consistently demonstrated its superior capability in sidelobe suppression and contrast enhancement. With only a modest computational overhead, the method is highly effective for both B-mode and Doppler imaging. Furthermore, its framework is readily adaptable to other advanced applications, including functional imaging and super-resolution imaging.

## V. CONCLUSION

In this study, we proposed a cross fusion and correlation beamformer to suppress the sidelobe artifacts and noise in RCA based ultrasound 3D imaging. Through comprehensive simulations and experimental validation, the proposed method demonstrates superior performance in both B-mode and Doppler flow imaging when compared to existing techniques. Specifically, it achieves the lowest sidelobe levels and the highest imaging contrast. These advancements suggest that the proposed beamformer has significant potential for clinical applications by enhancing the quality and reliability of 3D ultrasound imaging.